\documentclass[12pt,a4paper]{article}

\usepackage[export]{adjustbox}
\usepackage{dcolumn}
\usepackage{bm}
\usepackage{simplewick}
\usepackage{appendix}
\usepackage{relsize}
\usepackage[margin=1in]{geometry}
\usepackage{setspace}
\usepackage{amsmath}
\usepackage{amssymb}
\usepackage{amsfonts}
\usepackage[T1]{fontenc}
\usepackage{times}
\usepackage{graphicx}
\usepackage{mathtools}

\usepackage{setspace}
\usepackage{microtype}
\usepackage{siunitx}
\usepackage{nomencl}
\usepackage{fancyhdr}
\usepackage{xcolor}
\usepackage{booktabs}
\usepackage{multirow}
\usepackage{caption}
\usepackage{subcaption}
\usepackage{hyperref}
\usepackage{cleveref}
\usepackage{titletoc}
\usepackage{titlesec}
\usepackage{tocloft}
\usepackage{footmisc}
\usepackage{longtable}
\usepackage{bigints}
\usepackage{lscape}
\usepackage{array}
\usepackage{epstopdf}
\usepackage{glossaries}
\usepackage{pdfpages}
\DeclarePairedDelimiterXPP\BigOSI[2]%
  {\mathcal{O}}{(}{)}{}%
  {\SI{#1}{#2}}
\providecommand{\keywords}[1]
{
  \small	
  \textbf{\textit{Keywords---}} #1
}
\newpage

\title{Thermodynamics of Bardeen regular black hole with generalized uncertainty principle}
\author{Areeba Merriam$^{1}$, M. Zain Sarwar$^{2}$  \\
        \small $^{1}$Department of mathematics, Quaid-i-Azam University, Islamabad \\
        \small $^{2}$Department of physics, Quaid-i-Azam University, Islamabad \\
}
\date{\today}

\begin{document}

 \maketitle

\begin{abstract}
    
   This study explores the emission of massive charged spin-1 particles from the background of Bardeen regular spacetime by the semi-classical method used to study the Hawking radiation spectrum. We employed the Hamilton-Jacobi method and WKB approximation technique with the suitable form of the wave function to solve the Proca field equation. We calculated the tunneling probability of outgoing spin-1 particles and the corresponding thermodynamic temperature. Furthermore, we obtained the modified thermodynamic quantities like temperature, entropy as well as heat capacity by utilizing the quadratic form of generalized uncertainty principle (GUP) and minimal length. In the end, we investigated the local stability as well as phase transitions of the Bardeen black hole in the context of GUP-modified heat capacity.
    \end{abstract}
\tableofcontents

\section{Introduction}
The thermodynamics of gravitational objects is one of the most intriguing subjects today e.g the conception of temperature, pressure, volume, and entropy can be associated with them as well. Black holes can not emit radiations classically, while from the semi-classical perspective, Hawking was able to show that they can emit radiations \cite{hawking1975particle}. He studied the radiance of particles from black holes through the Wick Rotation method \cite{hawking1976black, hartle1976path}. He described it as a tunneling formalism happening due to vacuum fluctuations near the event horizon.
Hawking proclaimed theoretically that if we allow quantum effects then the black holes can emit radiations, known as the Hawking radiations. This was a considerably important discovery because not a single thing could go away from a black hole, classically. This showed that black holes have a well-defined thermodynamic temperature. \par 

Apart from Hawking's method of deriving the temperature of a black hole, now there have been established many ways to calculate the same. It can be derived from the black hole's surface gravity, the first law of thermodynamics, and the latest of them is the black hole tunneling method. The sense of the method is that the vacuum fluctuations (pair production) take place just inside the event horizon of a black hole. The positive energy particle materializes as an outgoing mode and the one with negative energy as an in-going mode. The one going outward is permitted to go from the horizon to infinity along the path which is classically forbidden. Hence the action of the particle turns complex and the tunneling probability will be determined by its imaginary part. However, on classical notes, a particle can always fall behind the horizon, so the action for in-going mode is always real.\par

Semi-classical modeling of the Hawking effect as a tunneling mechanism was first proposed in the 1990s which involves computing the imaginary part of the classical action for the proscribed trajectory of the emitted particle around the event horizon \cite{kraus1994simple}. This is equated to Boltzmann relation $\Gamma=\mathrm{e}^{-E/T}$ for emission at Hawking temperature, where $\Gamma$ is a probability measure, $E$ is energy and $T$ is temperature. Then by making use of the WKB (Wentzel, Kramers, Bril-louin) approximation technique the tunneling probability of the emitted particle arriving from inside to outside the event horizon can be written as
\begin{equation}
    \Gamma \propto \exp(-2  \operatorname {Im} \hat{\mathcal{S}}),
\end{equation}
where $\hat{\mathcal{S}}$ is the classical action of the path followed and $\operatorname {Im}\hat{\mathcal{S}}$ is an imaginary part of that action. We have two techniques to find out the imaginary part, one of them is the null geodesic method utilized by Parikh and Wilczek \cite{parikh2000hawking}, followed by the research of Kraus and Wilczek \cite{kraus1994simple,kraus1995self} and the other approach is Hamilton-Jacobi ansatz utilized by Angheben et al. \cite{angheben2005hawking}, which incorporates the method of complex path analysis by Srinivasan and Padmanabhan \cite{srinivasan1999particle,padmanabhan2004entropy}. Here, we will consider the latter one in this work. Related research has shown that this method can be widely applicable to the variety of spacetimes such as Schwarzschild \cite{chen2015vector}, charged fermions from Kerr-Newman \cite{kerner2008charged}, cylindrical black holes \cite{ahmed2011hawking,jamil2011hawking}, accelerating and rotating black holes \cite{gillani2011hawking,rehman2011charged,gillani2011tunneling}, and dynamical spacetimes \cite{di2008fermion,di2007hawking} etc. \par
Moreover, the study of GUP-corrections or quantum corrections is required in the semi-classical theory of black holes, since a black hole reduces its mass to the Planck length in consequence of the Hawking radiations. Though an absolute quantum gravitational approach is still lacking. The GUP model is one broadly accepted model to investigate the properties of quantum scale black holes. By employing this model we have performed the possible corrections for the thermodynamic quantities of the Bardeen black hole. It provides high-energy corrections regarding the theory of quantum gravity and the idea of minimal length. \par Such corrections have been studied for various classes of spacetimes such as for the case of general spherically symmetric metric \cite{mu2015minimal}, self dual black holes \cite{anacleto2015quantum}, Reissner-Nordström \cite{yoon2007entropy}, regular black holes \cite{maluf2018thermodynamics}, accelerating and rotating black holes \cite{rizwan2017gup}, and Hořava–Lifshitz gravity \cite{myung2009thermodynamics}. \par
The outline of this research is given below. In section (2) we have presented a short description of the Bardeen black hole. In section (3) the emission spectrum of charged vector particles from the background of a black hole spacetime is investigated. We worked out the tunneling amplitude of the outgoing particle and the expected Hawking temperature. In section (4) we have performed the quantum corrections for Hawking temperature, entropy as well as heat capacity by incorporating GUP. 

\section{Bardeen black hole}

There exist some solutions to Einstein's field equations which does not possess singularity at their origin i.e the curvature scalar does not goes to infinity. Bardeen presented such an exact solution in 1968 \cite{bardeen1968non} and after thirty years since its discovery, Ayon-Baeto and Garcia presented their interpretation in connection with non-linear electrodynamics \cite{ayon2000bardeen}. They proved that the associated charge acts as a magnetic monopole. Also, this spacetime satisfies the weak energy condition. Emission of spin-1/2 particles from Bardeen black hole has previously been studied in \cite{sharif2012charged}.\par
The line element of Bardeen regular black hole can be expressed as follows \cite{bardeen1981black}
\begin{equation}
    ds^2=- B(r) dt^2 + B(r)^{-1} dr^2 + r^2 d\theta^2 + r^2sin^2\theta d\phi^2,
\end{equation}
where 
\begin{equation}
    B(r)= \Bigg(1-\frac{2m(r)}{r}\Bigg)
\end{equation}
and
\begin{equation}\label{m}
    m(r)=M\Bigg(1+\bigg(\frac{q}{r}\bigg)^2\Bigg)^{-\frac{3}{2}},
\end{equation}
where $m$ is the mass and $q$ is the magnetic monopole charge of a black hole, respectively. It possess horizon at the location $r_+=2m(r_+)$. It reduces to Schwarzschild metric for $q=0$. The function $B(r)$ can be approximated asymptotically as
\begin{equation}
    B(r) \approx 1-\frac{2M}{r} +\frac{2Mq^2}{r^3}+ \mathcal{O}\Bigg(\frac{1}{r^5}\Bigg),
\end{equation}
which is not analogous to the Reissner–Nordström black hole. The event horizon $r_+$ of the spacetime can be derived by putting $B(r)=0$, as 
\begin{equation}
          r_+=\Bigg[\frac{1}{3}\Big(4M^2-3q^2\Big)+\frac{\sqrt[3]{2}}{3}\chi-\frac{24M^2q^2-16M^4}{3\sqrt[3]{2}\chi}\Bigg]^\frac{1}{2},
\end{equation}
where
\begin{equation}\label{chi}
    \chi= \Big(32M^6-72M^4q^2+27M^2q^4+3\sqrt{81M^4q^8-48M^6q^6}\Big)^{\frac{1}{3}}.
\end{equation}
From the above equation it can be interpreted that the condition $q<( 4\sqrt{3}/9) M$ must fulfil for the existence of inner and outer horizons. \par
Let us specify the following notations for simplicity
 \begin{gather*}
     C(r)=r^2,
     D(r)= \sin^2\theta.
 \end{gather*}
 The line element will become
 \begin{equation}
     ds^2= B(r) dt^2 + B(r)^{-1} dr^2 + C(r)d\theta^2+ C(r)D(r) d\phi^2.
 \end{equation}
The first law of thermodynamics for Bardeen spacetime is expressed as
\begin{equation}
    dm=TdS+\Phi dq,
\end{equation}
where $S$, $T$, and $\Phi$ are the entropy, temperature and electric potential of a   black hole, respectively. Also, the electric potential $A_0$ can be obtained from the above equation as \cite{sharif2010quantum}
\begin{equation}
    \Phi = A_0 =\frac{\partial m}{\partial q} \bigg|_{r=r_+}=\frac{3e}{2r_+^2}(r_+^2 +e^2)^{1/2}.
\end{equation}
\section{Emission of charged vector particles}
Now in order to look into the tunneling amplitude of spin-1 particles from the background of the Bardeen black hole, we used the Proca field for vector particles $\psi_\mu$ given by \cite{kruglov2014black, kruglov2014blackhole}
\begin{equation}
    D_\mu \psi^{\nu\mu} + \frac{m^2}{\hbar ^2} \psi^\nu =0.
\end{equation}
We have
\begin{equation}
    \psi_{\nu\mu} = D_\nu \psi_\mu-D_\mu \psi_\nu= \partial_\nu \psi_\mu - \partial_\mu \psi_\nu,
\end{equation}
where $\psi^\nu = (\psi^0, \psi^1, \psi^2, \psi^3)$, $D_\mu$ is covaraint derivative, and $\psi^{\nu\mu}$ is anti-symmetric tensor so we use the definition
\begin{equation}
    D_\mu \psi^{\nu\mu} =\frac{1}{\sqrt{-g}}\partial_\mu (\sqrt{-g}\psi^{\nu\mu}),
\end{equation}
 where $g$ is known as  the determinant of line element. Proca equation will have the following form
\begin{equation}
    \frac{1}{\sqrt{-g}}\partial_\mu (\sqrt{-g}\psi^{\nu\mu}) + \frac{m^2}{\hbar^2}\psi^\nu=0.
\end{equation}
The governing field equations for $\tilde{W}$-bosons is derived through the Lagrangian of the Glashow-Weinberg-Salam model \cite{li2015massive}. Let us recall the Lagrangian density surrounding the electromagnetic field 

\begin{equation}\label{L} %\label{c}
    \mathfrak{L} =-\frac{1}{2}\big(D^{+}_\mu \tilde{W}^{+}_\nu-D^{+}_\nu \tilde{W}^{+}_\mu\big)\big(D^{-\mu} \tilde{W}^{-\nu}-D^{-\nu} \tilde{W}^{-\mu}\big) + \frac{m^{2}}{\hbar^2}\tilde{W}^+_\mu \tilde{W}^{-\mu}-\frac{\iota}{\hbar}eF^{\mu\nu}\tilde{W}^+_\mu \tilde{W}^- _\nu,
\end{equation} 
where $D^{\pm}_\mu=\bar{\nabla}_\mu \pm \frac{\iota}{\hbar}eA_\mu$. $\bar{\nabla}_\mu$ represents the covariant derivative, $A_\mu=(A_0,0,0,0)$ is the electromagnetic potential of spacetime, $e$ represents the charge of boson, and $F^{\mu\nu} = \bar{\nabla}^\mu A^\nu - \bar{\nabla}^\nu A^\mu$ is electromagnetic field strength tensor. The required field equation for $\tilde{W}$-bosons can be obtained from the Lagrangian mentioned in Eq. (\ref{L}) as
\begin{equation}
     \begin{split}
        &\frac{1}{\sqrt{-g}}\partial_\mu\Big[\sqrt{-g}\big(D^{\pm\mu} \tilde{W}^{\pm\nu}-D^{\pm\nu} \tilde{W}^{\pm\mu}\big)\Big]+\frac{m^2}{\hbar^2}\tilde{W}^{\pm\nu} \pm \frac{\iota}{\hbar}eA_\mu \big(D^{\pm\mu} \tilde{W}^{\pm\nu}-D^{\pm\nu} \tilde{W}^{\pm\mu}\big)\\& \pm \frac{\iota}{\hbar}eF^{\nu\mu}\tilde{W}^{\pm}_\mu =0.
    \end{split}
\end{equation}
We can see that the field equation of the $\tilde{W}^+$ boson is the same as that of $\tilde{W}^-$ boson, so their tunneling processes should be similar too. So, we are going to study the motion of $\tilde{W}^+$ boson in detail. By using the relation $\bar{\nabla}_\mu \tilde{W}^+_\nu-\bar{\nabla}_\nu \tilde{W}^+_\mu=\partial_\mu \tilde{W}^+_\nu-\partial_\nu \tilde{W}^+_\mu$, the above equation of motion can be reformulated as
\begin{equation}
    \begin{split}
        &\frac{1}{\sqrt{-g}}\partial_\mu\Bigg[\sqrt{-g}g^{\mu\alpha}g^{\nu\beta}\Big(\partial_\beta \tilde{W}_\alpha^+-\partial_\alpha \tilde{W}_\beta^+ +\frac{\iota}{\hbar}eA_\beta \tilde{W}_\alpha^+-\frac{\iota}{\hbar}eA_\alpha \tilde{W}_\beta^{+}\Big)\Bigg] + \frac{\iota eA_\mu g^{\mu\alpha}g^{\nu\beta}}{\hbar}\\&\Big(\partial_\beta \tilde{W}_\alpha^+-\partial_\alpha \tilde{W}_\beta^+ +\frac{\iota}{\hbar}eA_\beta \tilde{W}_\alpha^+-\frac{\iota}{\hbar}eA_\alpha \tilde{W}_\beta^{+}\Big) + \frac{m^2 g^{\nu\beta}}{\hbar^2}\tilde{W}_\beta^+ +\frac{\iota}{\hbar}eF^{\nu\alpha}\tilde{W}_\alpha^+=0.
    \end{split}
\end{equation} \par
We will solve the above equation for $\nu=0,1,2,3$ and by summing over the other index. Now by making use of the WKB approximation, we can have the form of $\tilde{W}^+_\mu$ as given below
\begin{equation}\label{wkb}
    \tilde{W}^+_\mu=c_\mu \mathrm{exp}\Bigg[\frac{\iota}{\hbar}\hat{\mathcal{S}}(t,r,\theta,\phi)\Bigg],
\end{equation} 
where
\begin{equation}
    \hat{\mathcal{S}}(t,r,\theta,\phi)= \hat{\mathcal{S}_0}(t,r,\theta,\phi)+\hbar \hat{\mathcal{S}_1}(t,r,\theta,\phi)+\hbar^2\hat{\mathcal{S}_2}(t,r,\theta,\phi)+...
\end{equation}
We are having first order approximation here. For $\nu=0$ and $\mu=0,1,2,3$
\begin{equation}\label{my1}
    \begin{split}
        &c_0\bigg(-(\partial_r \hat{\mathcal{S}}_0)^2-\frac{(\partial_\theta \hat{\mathcal{S}}_0)^2}{BC}-\frac{(\partial_\phi \hat{\mathcal{S}}_0)^2}{BCD}-\frac{m^2}{B}\bigg)+c_1\bigg((\partial_r \hat{\mathcal{S}}_0)\big[\partial_t \hat{\mathcal{S}}_0+eA_t\big]\bigg) + c_2\bigg(\frac{(\partial_\theta \hat{\mathcal{S}}_0)}{BC}\\&\big[\partial_t \hat{\mathcal{S}}_0 +eA_t\big]\bigg) + c_3\bigg(\frac{(\partial_\phi \hat{\mathcal{S}}_0)}{BCD}\big[\partial_t \hat{\mathcal{S}}_0+eA_t\big]\bigg)=0.
    \end{split}
\end{equation}
For $\nu=1$ and $\mu=0,1,2,3$
\begin{equation}\label{my2}
    \begin{split}
        &c_0\bigg(-(\partial_r \hat{\mathcal{S}}_0)[eA_t+\partial_t \hat{\mathcal{S}}_0]\bigg)+c_1\bigg((\partial_t \hat{\mathcal{S}}_0+eA_t)^2-m^2B-\frac{B}{C}(\partial_\theta \hat{\mathcal{S}}_0)^2-\frac{B}{CD}(\partial_\phi \hat{\mathcal{S}}_0)^2\bigg)\\&+c_2\bigg(\frac{B}{C}(\partial_\theta \hat{\mathcal{S}}_0)(\partial_r \hat{\mathcal{S}}_0)\bigg)+c_3\bigg(\frac{B}{CD}(\partial_r \hat{\mathcal{S}}_0)(\partial_\phi \hat{\mathcal{S}}_0)\bigg)=0.
    \end{split}
\end{equation}
For $\nu=2$ and $\mu=0,1,2,3$
\begin{equation}\label{my3}
    \begin{split}
        &c_0\bigg(-\frac{(\partial_\theta \hat{\mathcal{S}}_0)}{B}\big[eA_t+\partial_t \hat{\mathcal{S}}_0\big]\bigg)+c_1\bigg(B(\partial_\theta \hat{\mathcal{S}}_0)(\partial_r \hat{\mathcal{S}}_0)\bigg)+c_2\bigg(\frac{(\partial_t \hat{\mathcal{S}}_0+eA_t)^2}{B}-m^2-B(\partial_r \hat{\mathcal{S}}_0)^2\\&-\frac{(\partial_\phi \hat{\mathcal{S}}_0)^2}{CD}\bigg)+c_3\bigg(\frac{(\partial_\theta \hat{\mathcal{S}}_0)(\partial_\phi \hat{\mathcal{S}}_0)}{CD}\bigg)=0.
    \end{split}
\end{equation}
And for $\nu=3$ and $\mu=0,1,2,3$
\begin{equation}\label{my4}
    \begin{split}
        &c_0\bigg(-\frac{(\partial_\phi \hat{\mathcal{S}}_0)}{B}\big[\partial_t \hat{\mathcal{S}}_0+eA_t\big]\bigg) + c_1\bigg(B(\partial_r \hat{\mathcal{S}}_0)(\partial_\phi \hat{\mathcal{S}}_0)\bigg) + c_2\bigg(\frac{(\partial_\theta \hat{\mathcal{S}}_0)(\partial_\phi \hat{\mathcal{S}}_0)}{C}\bigg) + c_3\bigg(\frac{(\partial_t \hat{\mathcal{S}}_0+eA_t)^2}{B}\\&-m^2-B(\partial_r \hat{\mathcal{S}}_0)^2-\frac{1}{C}(\partial_\theta \hat{\mathcal{S}}_0)^2\bigg)=0.
    \end{split}
\end{equation}
We employ the suitable ansatz of the following form by taking into account the spacetime symmetries
\begin{equation}
    \hat{\mathcal{S}_0}= -Et+\hat{W}(r)+\hat{J}(\theta,\phi)+I_0,
\end{equation}
From Eqs. (\ref{my1}-\ref{my4}) we arrive at the matrix representation
\begin{equation}
    \hat{\otimes}(c_0.c_1,c_2,c_3)^T=0,
\end{equation}
where $\hat{\otimes}$ is the $4\times 4$ matrix, the superscript $T$ specifies transformation to the transposed vector, its components are expressed as
\begin{align}
    \hat{\otimes}_{00}&= -\big(\hat{W}^\prime(r)\big)^2-\frac{\hat{J}_\theta^2}{BC}-\frac{\hat{J}_\phi^2}{BCD}-\frac{m^2}{B},& \hat{\otimes}_{01}&= \hat{W}^\prime(r)\big[eA_t-E\big],\nonumber\\ \hat{\otimes}_{02}&= \frac{\hat{J}_\theta}{BC}\big[eA_t-E\big],&\hat{\otimes}_{03}&=\frac{\hat{J}_\phi}{BCD} \big[eA_t-E\big],\nonumber\\ \hat{\otimes}_{10}&=-\hat{W}^\prime(r)\big[eA_t-E\big], & \hat{\otimes}_{11}&=\big[eA_t-E\big]^2-\frac{B \hat{J}_\theta^2}{C}-\frac{B \hat{J}_\phi^2}{CD}-m^2B,\nonumber\\ \hat{\otimes}_{12}&=\frac{\hat{W}^\prime(r)\hat{J}_\theta B}{C},& \hat{\otimes}_{13}&= \frac{\hat{W}^\prime(r)\hat{J}_\phi B}{CD},\nonumber\\ \hat{\otimes}_{20}&=-\frac{\hat{J}_\theta}{B}\big[eA_t-E\big],& \hat{\otimes}_{21}&= \hat{W}^\prime(r)\hat{J}_\theta B,\nonumber\\ \hat{\otimes}_{22}&=\frac{\big[eA_t-E\big]^2}{B}-\big(\hat{W}^\prime(r)\big)^2B-\frac{\hat{J}_\phi^2}{CD}-m^2,& \hat{\otimes}_{23}&= \frac{\hat{J}_\theta \hat{J}_\phi}{CD},\nonumber\\ \hat{\otimes}_{30}&=-\frac{\hat{J}_\phi}{B}\big[eA_t-E\big],& \hat{\otimes}_{31}&=\hat{W}^\prime(r) \hat{J}_\phi B,\nonumber\\ \hat{\otimes}_{32}&=\frac{\hat{J}_\theta \hat{J}_\phi}{C},& \hat{\otimes}_{33}&= \frac{\big[eA_t-E\big]^2}{B}-\big(\hat{W}^\prime(r)\big)^2B-\frac{\hat{J}_\theta}{C}-m^2.&
 \end{align}
where $\hat{W}^\prime=\partial_r \hat{\mathcal{S}}_0$, $\hat{J}_\theta=\partial_\theta \hat{\mathcal{S}}_0$, and $\hat{J}_\phi=\partial_\phi \hat{\mathcal{S}}_0$. We know that in order to get non-trivial solution of homogeneous system of linear equations we put determinant of the matrix $\hat{\otimes}$ equivalent to zero, i.e, det$\hat{\otimes}=0$. So we get
 \begin{equation}
     m^2\Big[B(D \hat{J}_\theta^2+\hat{J}_\phi^2)+CD\Big(-\big[E-e A_t\big]^2+B\big(m^2+B(\hat{W}^\prime(r))^2\big)\Big)\Big]^3=0.
 \end{equation}
 By simplifying for the radial part we obtain 
 \begin{equation}
    \big(\hat{W}^\prime(r)\big)^2=\frac{\big[E-eA_t\big]^2-B\Big(m^2+\frac{\hat{J}_\theta^2}{C}+\frac{\hat{J}_\phi^2}{CD}\Big)}{B^2},
\end{equation}
\begin{equation}\label{pole}
\begin{split}
     \hat{W}_{\pm}=&\pm \bigintsss \frac{dr}{B(r)} \sqrt{\big[E-eA_t\big]^2-B(r) \Bigg(\frac{\hat{J}_\theta^2}{C}+\frac{\hat{J}_\phi^2}{CD }+m^2}\Bigg),\\&
     \pm \bigintsss \frac{dr}{\frac{1-2m(r)}{r}} \sqrt{\big[E-eA_t\big]^2-\frac{1-2m(r)}{r} \Bigg(\frac{\hat{J}_\theta^2}{C}+\frac{\hat{J}_\phi^2}{CD }+m^2}\Bigg).
\end{split}
\end{equation}

Here, $\hat{W}_+$ is the solution that relates to the vector particles traveling away from the event horizon (i.e outgoing) while $\hat{W}_-$ relates to the particles traveling towards the horizon (i.e incoming).
The probabilities of crossing the event horizon as emission or absorption are given as
\begin{equation}\label{em}
    \mathcal{P}_{[\mathrm{emission}]} \propto \exp\Bigg[\frac{-2}{\hbar}\operatorname {Im}\hat{\mathcal{S}}_0\Bigg]=\exp\Bigg[\frac{-2}{\hbar}(\operatorname {Im}\hat{W}_+ + \operatorname {Im}I_0)\Bigg],
\end{equation}
\begin{equation}\label{abs}
    \mathcal{P}_{[\mathrm{absorption}]} \propto \exp\Bigg[\frac{-2}{\hbar}\operatorname {Im}\hat{\mathcal{S}}_0\Bigg]=\exp\Bigg[\frac{-2}{\hbar}(\operatorname {Im}\hat{W}_- + \operatorname {Im}I_0)\Bigg].
\end{equation}
So, the tunneling rate of outgoing particle from Bardeen black hole can be obtained as
\begin{equation}
   \Gamma=\frac{\mathcal{P}_{[\mathrm{emission}]}}{\mathcal{P}_{[\mathrm{absorption}]}}= \frac{\exp\big[\frac{-2}{\hbar}(\operatorname {Im}\hat{W}_+ + \operatorname {Im}I_0)\big]}{\exp\big[\frac{-2}{\hbar}(\operatorname {Im}\hat{W}_- + \operatorname {Im}I_0)\big]}.
\end{equation}
The incoming vector particle outside the event horizon should have $100\%$ feasibility of entering inside the black hole i.e. $\mathcal{P}_{[\mathrm{absorption}]}=1$, we know that $\hat{W}_+=-\hat{W}_-$ so we have $\operatorname {Im}I_0=-\operatorname {Im}\hat{W}_-$. The tunneling probability of a vector particle is given as
\begin{equation}\label{prob}
    \Gamma=\exp \Bigg[\frac{-4}{\hbar}\operatorname {Im}\hat{W}_+\Bigg].
\end{equation}
Now we expand $B(r)$ in Taylor series, we get
\begin{equation}
    B(r)= B(r_+)+B^\prime (r_+)(r-r_+).
\end{equation}
$B^\prime(r_+)$ can be calculated as
\begin{equation}
    B^\prime (r_+)=\frac{2Mr_+(r_+^2-2q^2)}{(r_+^2+q^2)^{5/2}}.
\end{equation}
In Eq. (\ref{pole}) there is a simple pole at the event horizon $r_+= 2m(r_+)$ so we apply the residue theory to solve the complex integral
\begin{equation}
    \hat{W}_{\pm}=\pm\frac{[E-eA_t]}{B^\prime (r_+)}\int \frac{dr}{(r-r_+)}.
\end{equation}
So we arrive at the following solution
\begin{equation}
    \hat{W}_{\pm}=\pm\frac{\pi\iota[E-eA_t]}{B^\prime (r_+)}.
\end{equation}
In an explicit form it is written as
\begin{equation}
    \hat{W}_{\pm}=\pm\frac{\pi\iota[E-eA_t](r_+^2+q^2)^{5/2}}{2Mr_+(r_+^2-2q^2)},
\end{equation}
\begin{equation}
    \mathrm{Im}\hat{W}_{\pm}=\pm \frac{\pi[E-eA_t](r_+^2+q^2)^{5/2}}{2Mr_+(r_+^2-2q^2)} =\pm \frac{\pi[E-eA_t]}{B^\prime(r_+)}.
\end{equation}
Now the tunneling probability of vector particles emitting away from the horizon $r_+$ is given by imaginary part of the radial term. Setting $\hbar$ to unity, we get
\begin{equation}
    \Gamma= \mathrm{exp}\Bigg[-\frac{4}{\hbar}\mathrm{Im}\hat{W}_+\Bigg]= \mathrm{exp}\Bigg[-\frac{4\pi[E-eA_t](r_+^2+q^2)^{5/2}}{2Mr_+(r_+^2-2q^2)}\Bigg].
\end{equation}
By comparing the above equation with the Boltzmann expression for emission $\Gamma=e^{-E/T}$, we can obtain the Hawking temperature for the Bardeen regular spacetime in consistence with \cite{akbar2012thermodynamics}
\begin{equation}\label{temp}
    T_H=\frac{B^\prime(r_+)}{4\pi}= \frac{Mr_+(r_+^2-2q^2)}{2\pi(r_+^2+q^2)^{5/2}}. 
\end{equation}
The mass $M$ of a black hole in the horizon radius terms can be specified from Eq. (\ref{m}) as
\begin{equation}
M=\frac{(r_+^2 + q^2)^\frac{3}{2}}{2 r_+2},
\end{equation}
hence Eq. (\ref{temp}) will become
\begin{equation}
    T_H= \frac{(r_+^2-2q^2)}{4\pi r_+(r_+^2+q^2)}.
\end{equation}
\section{Modifications in thermodynamics by GUP}
In the following part, we show the consequences of GUP on the thermodynamic quantities of Bardeen regular black hole based on the recent works in refs. \cite{nouicer2007quantum, anacleto2015quantum, casadio2018generalised, maluf2018thermodynamics}. Lets us start from the expression of GUP with the quadratic term in momentum, given as
\begin{equation}
    \Delta x \Delta p \geq \hbar \Bigg(1-\beta \Delta p+\beta^2(\Delta p)^2\Bigg),
\end{equation}
where $\Delta p$ and $\Delta x$ are the uncertainties in momentum and position, respectively while $\beta$ is the dimensionless parameter $(\beta=\lambda l_p/\hbar)$. \par We can revise the above equation as
\begin{equation}
    \Delta x \Delta p \geq \hbar \Bigg(1-\frac{\lambda l_p}{\hbar}\Delta p+\frac{\lambda^2 l_p^2}{\hbar}(\Delta p)^2\Bigg),
\end{equation}
where $\lambda$ is dimensionless quantum gravity parameter, $l_p$ is Planck's length, which is equals to $l_p=\sqrt{\hbar G/c^3}=M_pG/c^2 \approx 10^{-35}m$, and $M_p=\sqrt{\hbar c/G}$ is Planck's mass. Here the correction terms in the usual Heisenberg's relation are due to the gravitational effects. \par The GUP expression can be rewritten as
\begin{equation}
    \Delta p \geq \frac{\hbar (\Delta x+\lambda l_p)}{2\lambda^2l_p^2}\Bigg(1-\sqrt{1-\frac{4\lambda^2l_p^2}{(\Delta x+\lambda l_p)^2}}\Bigg),
\end{equation}
where we have chosen the negative solution. Since $l_p/\Delta x << 1$, we use the Taylor series such that
\begin{equation}
    \Delta p \geq \frac{1}{2\Delta x}\Bigg(1-\frac{\lambda}{2(\Delta x)}+\frac{\lambda^2}{2(\Delta x)^2}+...\Bigg),
\end{equation}
 here we have selected the units $G=c=k_b=\hbar=l_p=1$. For $\lambda=0$ the above equation reduces to the original Heisenberg's uncertainty relation i.e. $\Delta x \Delta p \geq 1$, where the factor of half has been absorbed in $\Delta x$. 
From the standard dispersion relation $E^2=p^2+m^2$, we can have energy bound for the black hole $E\Delta x \geq 1$. Consequently, the energy correction up to second order in Planck length reads as
\begin{equation}
    E_{\mathrm{GUP}} \geq E\Bigg(1-\frac{\lambda}{2(\Delta x)}+\frac{\lambda^2}{2(\Delta x)^2}+...\Bigg),
\end{equation}
where $E_{GUP}$ is the corrected energy. The tunneling probability will be modified for vector particles crossing the event horizon with the GUP corrected energy
\begin{equation}
    \Gamma \simeq \mathrm{exp}\Bigg[-\frac{4\pi E_{\mathrm{GUP}}}{B^\prime(r_+)}\Bigg]= \mathrm{exp}\Bigg[-\frac{E_{\mathrm{GUP}}}{T_H}\Bigg].
\end{equation}
On comparison with the Boltzmann expression, we can have the GUP-modified temperature as
\begin{equation}
    T_{\mathrm{GUP}}= T_H\Bigg(1-\frac{\lambda}{2(\Delta x)}+\frac{\lambda^2}{2(\Delta x)^2}\Bigg)^{-1},
\end{equation}
where $T_H$ is given in (\ref{temp}). As we are approximating our solution near the horizon so we can choose the minimum uncertainty in the position of outgoing particle is of the order of horizon radius of the black  hole spacetime as 
\begin{equation}
\Delta x = 2r_+.
\end{equation}
That choice of the length scale has been argued of other configurations as well, such as in refs. \cite{kuchiev2004scattering, medved2004conceptual}.
Hence the above equation takes the form
\begin{equation}
\begin{split}
T_{\mathrm{GUP}} &\leq  T_H \Bigg(1-\frac{\lambda}{4r_+}+\frac{\lambda^2}{8r_+^2}\Bigg)^{-1},\\&
    =  \Bigg(\frac{(r_+^2-2q^2)}{4\pi r_+(r_+^2+q^2)}\Bigg) \Bigg(1-\frac{\lambda}{4r_+}+\frac{\lambda^2}{8r_+^2}\Bigg)^{-1}.
\end{split}
\end{equation}
 
 \begin{figure}[htp] \centering{
\includegraphics[scale=0.5]{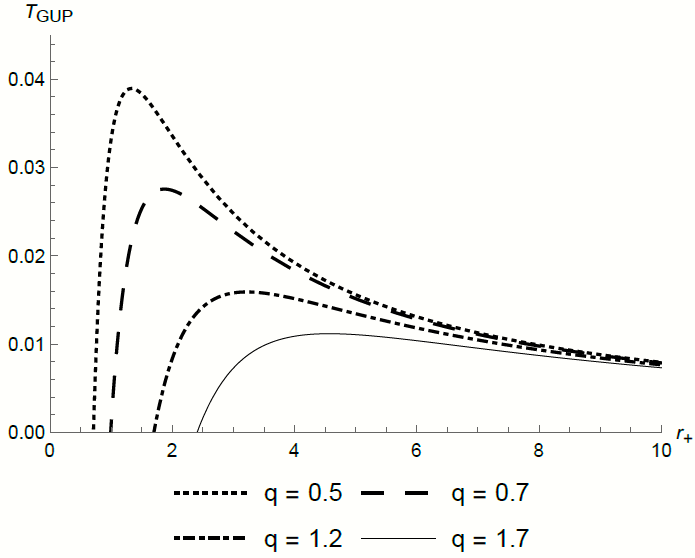}
}
\caption{Plot between corrected temperature and horizon radius for different values of $q$ and fixed value of $\lambda=3$.}
\label{tr1}
\end{figure} 
\begin{figure}[htp] \centering{
\includegraphics[scale=0.6]{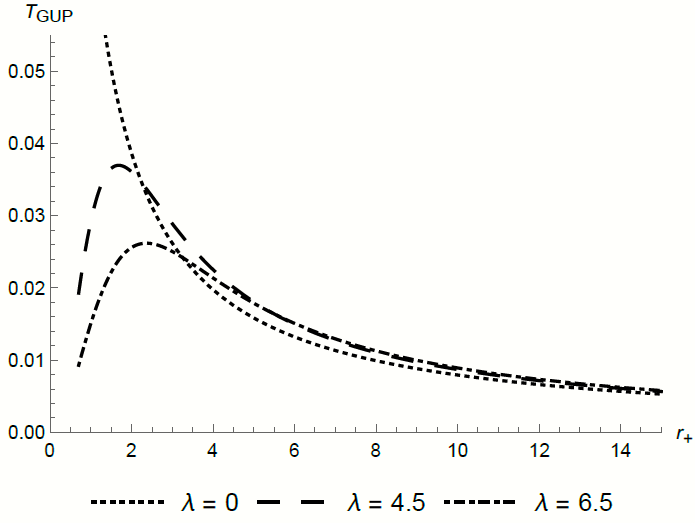}
}
\caption{Plot between corrected temperature and horizon radius with varying quantum gravity parameter and fixed value of $q=0.2$.}
\label{tr2}
\end{figure}
In Fig. (\ref{tr1}) we can observe a rapid spike in corrected temperature for a small radius. After attaining a certain height,  $T_{\mathrm{GUP}}$ starts to decrease gradually with increasing horizon radius. Also, note that $T_{\mathrm{GUP}}$ decreases as $q$ increases. This decreasing, as well as positive  temperature  with respect to the horizon radius, indicates the thermodynamic stability of the Bardeen regular black hole. \par
On the other hand, Fig. (\ref{tr2}) illustrates the case with both standard Hawking temperature $(\lambda=0)$ and modified Hawking temperature ($\lambda=4.5$ and $\lambda=6.5$). Upon comparison, we can see that the quantum gravity effects are negligible on a larger scale with increasing horizon radius. However, as the horizon radius of a black hole start to decrease and almost reaches Planck length the effects become considerably prominent. As per Fig., the standard Hawking temperature increases in a monotonic way with decreasing radius. However, the corrected Hawking temperature will tend to decrease abruptly upon reaching a certain height and eventually arrive at zero. This result suggests the existence of a black hole remnant. The temperature becomes nonphysical upon further decreasing horizon radius that depicts the unstable state of a black hole. \par 
Now by incorporating the first law of black hole thermodynamics, we can obtain the GUP-modified entropy as
\begin{equation}
    S_{\mathrm{GUP}}= \bigintssss \frac{d m}{T_{\mathrm{GUP}}}= \bigintssss \frac{\kappa dA}{8\pi T_{\mathrm{GUP}}},
\end{equation}
where $A=4S=4\pi r^2$ is the area of black hole. Substituting $\kappa=2\pi T_H$ and $dA=8\pi r_+ dr_+$ in the above equation, we get
\begin{equation}\label{entr}
\begin{split}
    S_\mathrm{GUP}=& \bigintssss \frac{2\pi r_+ T_H}{T_\mathrm{GUP}}dr_+ = \bigintssss 2\pi r_+ \Bigg(1-\frac{\lambda }{4r_+}+\frac{\lambda^2}{8 r_+^2}+...\Bigg)dr_+, \\&
    =\Bigg(\pi r_+^2 -\frac{\lambda \pi r_+}{2} +\frac{\lambda^2\pi \mathrm{ln} r_+}{4}\Bigg).
    \end{split}
\end{equation}
The first term is the standard black hole entropy while others represent the correction terms due to $\lambda$-parameter. In terms of surface area of the horizon the above equation can be rewritten as
\begin{equation}
    S_{\mathrm{GUP}}=\frac{A}{4}-\frac{\lambda}{4}\sqrt{A\pi}+\frac{\lambda^2\pi}{8} \mathrm{ln}\frac{A}{4}.
\end{equation}
Hence we arrive at the logarithmic corrections to entropy owing to GUP effects as indicated in \cite{kaul2000logarithmic}. The recent studies show us that the upper bound for gravitational parameter $\lambda$ is approximately $10^{10}$ \cite{scardigli2015gravitational}. Also, note that for $\lambda=0$, the well-known expression for the semi-classical Bekenstein-Hawking entropy is recovered.
\begin{figure}[htp] \centering{
\includegraphics[scale=0.5]{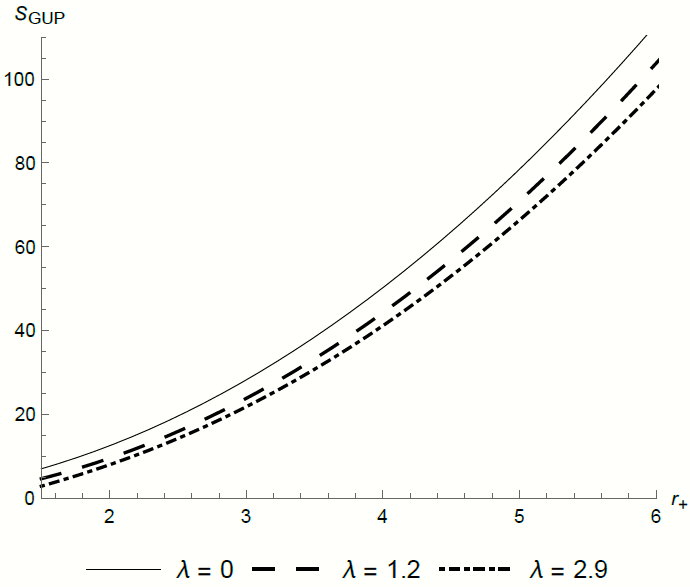}
}
\caption{Plot between corrected entropy and horizon radius with varying quantum gravity parameter.}
\label{entropy}
\end{figure}
The corrected black hole entropy as a function of $r_+$ is shown in Fig. (\ref{entropy}). We can see that the entropy is considerably increasing with the increase in horizon radius. Moreover, it is somehow decreasing with increase in correction parameter. \par
Now we will work out the GUP-corrected heat capacity for Bardeen black hole at horizon $r_+$, analogous to the classical case. At constant charge $q$, the corrected heat capacity can be expressed by the following relation
\begin{equation}
    C_\mathrm{GUP}\Big|_q=\frac{\partial M}{\partial T_\mathrm{GUP}}\Big|_q =\frac{\partial M}{\partial r_+}.\frac{\partial r_+}{\partial T_\mathrm{GUP}}.
\end{equation}
The partial derivatives of $M$ and $T_\mathrm{GUP}$ are given below, respectively, as
\begin{equation}
    \frac{\partial M}{\partial r_+}=\frac{(-2q^2+r_+^2)(q^2+r_+^2)^\frac{1}{2}}{2r_+^3},
\end{equation}
 \begin{equation}
     \frac{\partial T_\mathrm{GUP}}{\partial r_+}=\frac{2(-12\lambda q^2 r_+^3-8r_+^2(-2q^4-7q^2r_+^2+r_+^4)+\lambda^2(-2q^4+5q^2r_+^2+r_+^4))}{\pi (q^2+r_+^2)^2(\lambda^2 -2\lambda r_+ +8r_+^2)^2}.
 \end{equation}
 By making use of the above mentioned equations, the corrected specific heat capacity at constant $q$ reads as
 \begin{equation} 
    C_\mathrm{GUP}\Big|_q = \frac{\pi (-2q^2 + r_+^2)(q^2+r_+^2)^{5/2}(\lambda^2 -2\lambda r_+ +8r_+^2)^2}{4r_+^3(-12\lambda q^2 r_+^3 -8 r_+^2(-2q^4-7q^2 r_+^2 +r_+^4)+\lambda^2(-2q^4+5q^2 r_+^2+r_+^4))}.
 \end{equation}
 The above equation tells us that the Bardeen regular black hole undergoes both the first as well as second-order phase transitions in the presence of the GUP parameter. Discontinuities in the graph show the second-order phase transition while the shift from negative to positive values of heat capacity at a smaller radius shows the first-order phase transition. \par
\begin{figure}[htp]  \centering{
\includegraphics[scale=0.6]{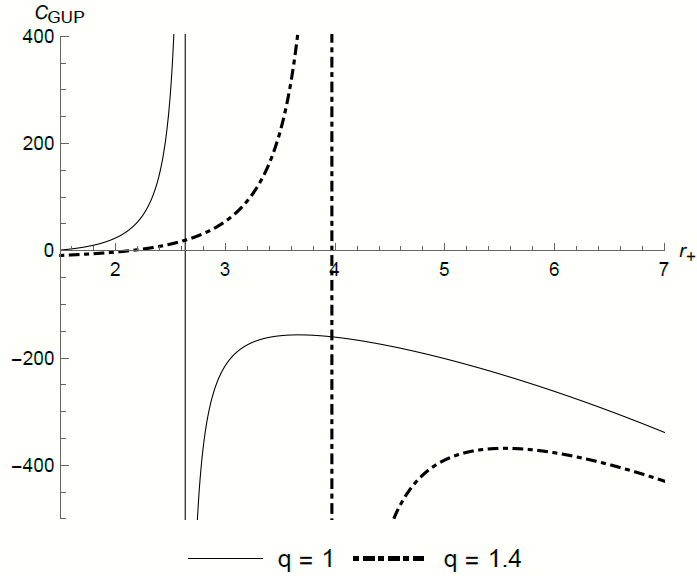}
}
\caption{Plot between heat capacity at constant charge versus horizon radius is given. Positive values of specific heat in the graph show stable regions while discontinuities show phase transitions. Here, $\lambda$ is set to be 1 and $q$ is varied.}
\label{heat}
\end{figure}

  As illustrated in Fig. (\ref{heat}), heat capacity is negative for a very small radius, positive for in-between and then again negative for the larger radius of a black hole spacetime. In this way, we obtain the local stability regions for Bardeen black hole.
 Now if we put $\lambda=0$ in the above equation, the usual heat capacity for Bardeen black hole is recovered as given in \cite{akbar2012thermodynamics}
 \begin{equation}
     \lim_{\lambda \to 0}C_\mathrm{GUP}\Big|_q = C\Big|_q =\frac{2\pi (r_+^2 + q^2)^{5/2} (r_+^2 -2q^2)}{r_+ (2q^4 +7q^2 r_+^2 -r_+^4)}.
 \end{equation}

\section{Concluding remarks}\label{SC}
In this article, we have investigated the black hole tunneling formalism to study the emission of charged vector bosons from the horizon of Bardeen regular spacetime. We employed the Hamilton-Jacobi technique as well as WKB approximation to investigate the Hawking temperature. The resulted Hawking temperature is found to be consistent with the earlier results. Hence, the result provided the evidence up to remarkable precision for the generalization of black hole radiance. \par Moreover, GUP was introduced from the energy correction of an emitted particle due to the effects of gravity around the horizon of a black hole. By considering the quadratic form of GUP, we derived the modified Hawking temperature and entropy. The GUP parameter provided with the corrected  logarithmic term to the entropy. The graphical interpretation of modified entropy tells us that it increases with the increase in horizon radius $r_+$. Moreover, it behaves abnormally for a small range near the center of the black hole by maintaining the negative value. Then from the modified results of Hawking temperature, we obtained heat capacity at constant charge $q$, which indicated the local stability as well as thermodynamic phase transitions of Bardeen regular black hole. Also, by assuming the quantum gravity parameter as zero i.e for $\lambda=0$, the standard heat capacity of the Bardeen black hole is recovered.

\bibliographystyle{elsarticle-num}
\bibliography{main.bib}

\end{document}